\documentclass[aip,reprint, jmp, amsmath, amssymb]{revtex4-1}
\usepackage{graphicx}
\usepackage{dcolumn}
\usepackage{amsmath}
\usepackage{bm}

\begin{document}

\title{Identifying graphene layers via spin Hall effect of light}

\author{Xinxing Zhou}
\author{Xiaohui Ling}
\author{Hailu Luo}
\email{hailuluo@hnu.edu.cn}
\author{Shuangchun Wen}
\email{scwen@hnu.edu.cn}
\affiliation{Key Laboratory for
Micro-/Nano-Optoelectronic Devices of Ministry of Education, College
of Physics and Microelectronic Science, Hunan University, Changsha
410082, People's Republic of China}
\date{\today}

\begin{abstract}
The spin Hall effect (SHE) of light is a useful metrological tool
for characterizing the structure parameters variations of
nanostructure. In this letter we propose using the SHE of light to
identify the graphene layers. This technique is based on the
mechanism that the transverse displacements in SHE of light are
sensitive to the variations of graphene layer numbers.
\end{abstract}

\maketitle

The quick and convenient technique for identifying the layer numbers
of graphene film is important for accelerating the study and
exploration of graphene material~\cite{Geim2009}. There have many
methods for determining the layer numbers of graphene film, yet
existing limitation. For instance, atomic force microscopy technique
is the straight way to determine the layer numbers of graphene. But
this method shows a slow throughput and may induce damage to the
sample. Unconventional quantum Hall effects~\cite{Zhang2005} are
usually used to distinguish one layer and two layers graphene from
multiple layers. Raman spectroscopy~\cite{Gupta2006} shows
characteristic for quick and nondestructive measuring the layer
numbers of graphene. However, it is not obvious to tell the
differences between bilayer and a few layers of graphene
films~\cite{Ni2007}.

The spin Hall effect (SHE) of light appears as a transverse
spin-dependent splitting, when a spatially confined light beam
passes from one material to another with different refractive
index~\cite{Onoda2004,Bliokh2006,Gorodetski2008,Luo2011a,Bliokh2008,Hosten2008,Qin2009,Aiello2008,Hermosa2011,Menard2009}.
The SHE of light is the photonic version of the spin Hall effect in
electronic systems, in which the spin photons play the role of the
spin charges, and a refractive index gradient plays the role of the
electric potential gradient. The SHE of light holds great potential
applications, such as manipulating electron spin states and
precision metrology. Importantly, the SHE of light can serve as a
useful metrological tool for characterizing the structure parameters
variations of nanostructure due to their sensitive
dependence~\cite{Hosten2008}. For example, in the previous work, we
have measured the thickness of the nanometal film via weak
measurements~\cite{Zhou2012a}. Therefore the SHE of light may have a
potential to determine the layer numbers of graphene.
\begin{figure}
\includegraphics[width=9cm]{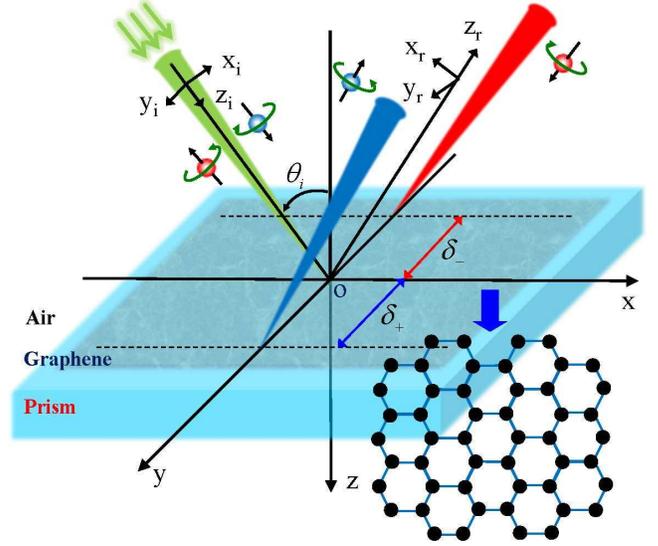}
\caption{\label{Fig1} (Color online) Schematic of SHE of light on a
graphene film. A linearly polarized beam reflects on the model and
then splits into left- and right-circularly polarized light,
respectively. $\delta_{+}$ and $\delta_{-}$ indicate the transverse
shift of left- and right-circularly polarized components. Here,
$\theta_{i}$ is the incident angle and the inset shows the atomic
structure of graphene. }
\end{figure}

In this letter, we propose a simple method for measuring the layer
numbers of graphene. We find that SHE of light can serve as an
advantageous metrology tool for characterizing the layer numbers of
graphene. The rest of the paper is organized as follows. First a
general propagation model is used to analyze the SHE of light on a
graphene film. We establish the relationship between the transverse
shifts and the graphene layers. Then, we focus our attention on the
experiment for judging the layer numbers of graphene. Here we use
the SHE of light to choose the suitable refractive index of graphene
obtaining from the literature. And then, with the suitable
refractive index, the layer numbers of an unknown graphene film can
be detected. We introduce the weak measurements technique and the
sample is a BK7 substrate transferred with the graphene film using
the chemical vapor deposited (CVD) method. Finally, we summarize the
main results of the paper.

We first theoretically analyze the SHE of light on graphene film and
establish the relationship between transverse shifts and the layer
numbers of graphene. Figure~\ref{Fig1} schematically illustrates the
SHE of light reflection on a graphene film in Cartesian coordinate
system. The $z$ axis of the laboratory Cartesian frame ($x,y,z$) is
normal to the interface of the graphene film at $z=0$. The incident
and reflected electric fields are presented in coordinate frames
($x_i,y_i,z_i$) and ($x_r,y_r,z_r$), respectively. In the spin basis
set, the angular spectrum can be written as
$\tilde{\mathbf{E}}_i^H=(\tilde{\mathbf{E}}_{i+}+\tilde{\mathbf{E}}_{i-})/{\sqrt{2}}$
and
$\tilde{\mathbf{E}}_i^V=i(\tilde{\mathbf{E}}_{i-}-\tilde{\mathbf{E}}_{i+})/{\sqrt{2}}$.
Here, $H$ and $V$ represent horizontal and vertical polarizations,
respectively. The positive and negative signs represent the left-
and right-circularly polarized (spin) components, respectively.

The incident monochromatic Gaussian beam can be formulated as a
localized wave packet whose spectrum is arbitrarily narrow, and can
be written as
\begin{equation}
\widetilde{\mathbf{E}}_{i\pm}=(\mathbf{e}_{ix}+i\sigma\mathbf{e}_{iy})\frac{w_{0}}{\sqrt{2\pi}}\exp
\left[-\frac{w_{0}^{2}(k_{ix}^{2}+k_{iy}^{2})}{4}\right]\label{3},
\end{equation}
where $w_{0}$ is the beam waist. The polarization operator
$\sigma=\pm1$ corresponds to left- and right-circularly polarized
light, respectively. In this work, we only consider the incident
light beam with horizontal polarization and vertical polarization
can be analyzed in the similar way. Using the reflection
matrix~\cite{Luo2011b,Zhou2012b}, we can obtain the expressions of
the reflected angular spectrum
\begin{equation}
\widetilde{\mathbf{E}}_{r}=\frac{r_{p}}{\sqrt{2}}\left[\exp(+ik_{ry}\delta_{r})\widetilde{\mathbf{E}}_{r+}+\exp(-ik_{ry}\delta_{r})\widetilde{\mathbf{E}}_{r-}\right]\label{5}.
\end{equation}
Here, $\delta_{r}=(1+r_{s}/r_{p})\cot\theta_{i}/k_{0}$, $r_{p}$ and
$r_{s}$ denote Fresnel reflection coefficients for parallel and
perpendicular polarizations, respectively. $k_{0}$ is the wave
number in free space. And the $\widetilde{\mathbf{E}}_{r\pm}$ can be
written as
\begin{equation}
\widetilde{\mathbf{E}}_{r\pm}=(\mathbf{e}_{rx}+i\sigma\mathbf{e}_{ry})\frac{w_{0}}{\sqrt{2\pi}}\exp
\left[-\frac{w_{0}^{2}(k_{rx}^{2}+k_{ry}^{2})}{4}\right]\label{7}.
\end{equation}

At any given plane $z_r=const.$, the transverse displacement of
field centroid compared to the geometrical-optics prediction is
given by
\begin{equation}
\delta_{\pm}= \frac{\int\int \tilde{\mathbf{\xi}}^{\ast}_{r\pm}
i\partial_{k_{ry}}\tilde{\mathbf{\xi}}_{r\pm} dk_{rx}
dk_{ry}}{\int\int
\tilde{\mathbf{\xi}}^{\ast}_{r\pm}\tilde{\mathbf{\xi}}_{r\pm}
dk_{rx} dk_{ry}}\label{centroid},
\end{equation}
where $\tilde{\mathbf{\xi}_{r}}_{\pm}$ = $r_p\exp(\pm
ik_{ry}\delta_{r})\widetilde{\mathbf{E}}_{r\pm}$. Calculating the
reflected displacements of the SHE of light requires the explicit
solution of the boundary conditions at the interfaces. Thus, we need
to know the generalized Fresnel reflection of the graphene film,
\begin{eqnarray}
r_{A}=\frac{R_{A}+R_{A}^{'}\exp(2ik_{0}\sqrt{n^{2}-\sin^{2}\theta_{i}}d)}{1+R_{A}R_{A}^{'}\exp(2ik_{0}\sqrt{n^{2}-\sin^{2}\theta_{i}}d)}.
\end{eqnarray}
Here, $A\in\{p,s\}$, $R_{A}$ and $R_{A}^{'}$ is the Fresnel
reflection coefficients at the first interface and second interface,
respectively. $n$ and $d$ represent the refractive index and
thickness of the graphene film, respectively. The thickness of the
graphene film is about $d=m\Delta d$ in which $m$ denotes the layer
numbers and $\Delta d$ represents the thickness of single layer
graphene ($\Delta d$ is about $0.34$ nm). So far we have established
the relationship between the transverse shifts and the graphene
layers. After obtaining the displacements of SHE of light, we can
determine the layer numbers of graphene.
\begin{figure}
\includegraphics[width=9cm]{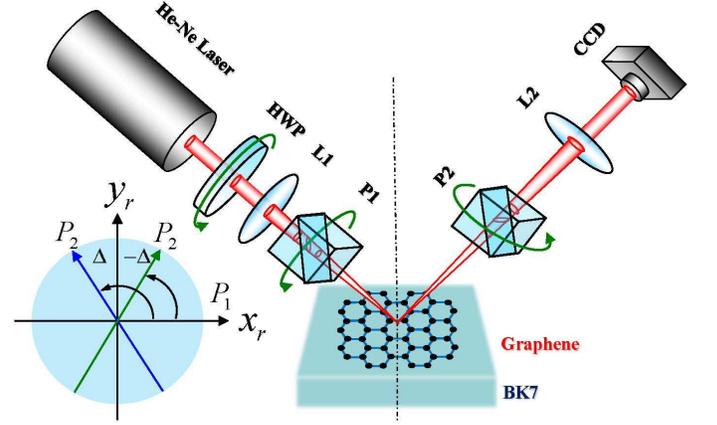}
\caption{\label{Fig2} (Color online) Experimental setup: The sample
is a BK7 glass transferred with the graphene film. L1 and L2, lenses
with effective focal length $50$ mm and $250$ mm, respectively. HWP,
half-wave plate (for adjusting the intensity). P1 and P2, Glan Laser
polarizers. CCD, charge-coupled device (Coherent LaserCam HR). The
light source is a $17$ mW linearly polarized He-Ne laser at $633$ nm
(Thorlabs HRP170). The inset shows that the angle between P1 and P2
is $90^{\circ}\pm\Delta$. }
\end{figure}

\begin{figure}
\includegraphics[width=6cm]{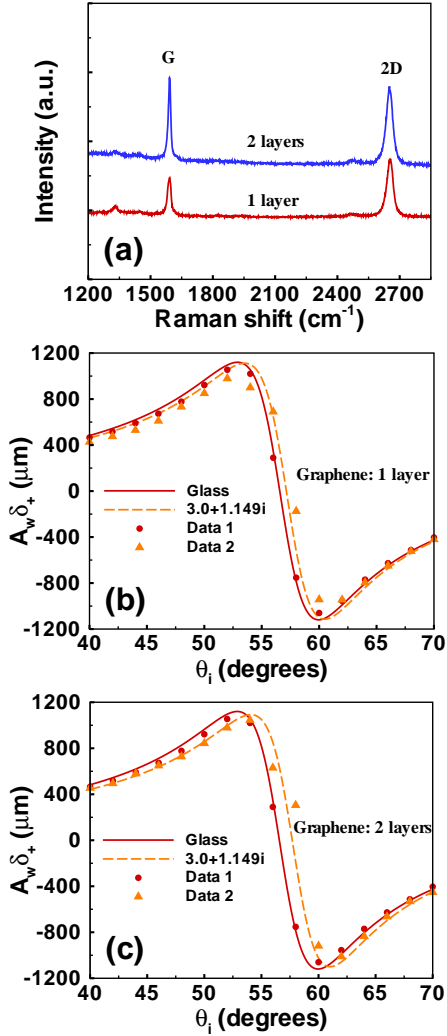}
\caption{\label{Fig3} (Color online) Raman spectra of the samples
and the graphene refractive index selection in the case of
horizontal polarization. (a) Raman spectra of one, two graphene
layers. (b) represents the transverse displacements under the
condition of single layer graphene. We choose the thickness of one
layer graphene film as 0.34 nm. The transverse shifts in the case of
two layers graphene film are shown in (c). Here, the lines represent
the theoretical results. The circle and triangle show the
experimental results obtained from the air-prism and different
graphene-prism condition via weak measurements. The refractive index
of the BK7 substrate is chosen as $n=1.515$ at $633$ nm. }
\end{figure}

In this work a signal enhancement technique known as the weak
measurements~\cite{Aharonov1988,Ritchie1991} is used to measure the
tiny transverse displacements. The experimental setup shown in
Fig.~\ref{Fig2} is similar to that in Refs~\cite{Luo2011a,Qin2009}.
A Gaussian beam generated by a He-Ne laser passes through a short
focal length lens (L1) and a polarizer (P1) to produce an initially
horizontal polarization beam. Here the half-wave plate (HWP) is used
to control the light intensity. When the beam impinges onto the
graphene-prism interface, the SHE of light takes place, manifesting
itself as the opposite displacements of the two spin components.
Then the two components interfere destructively after the second
polarizer (P2) which is oblique to P1 with an angle of
$90^{\circ}\pm\Delta$. We note that the incident light beam is
preselected in the H polarization state by P1 (along to the
$x_{i}$-axis) and then postselected by P2 in the polarization state
with
$\mathbf{V}=\sin\Delta\mathbf{e}_{rx}+\cos\Delta\mathbf{e}_{ry}$. In
this condition, we choose the angle $\Delta=2^{\circ}$. Then we use
L2 to collimate the beam and make the beam shifts insensitive to the
distance between L2 and the CCD. The reflected field at the plane of
$z_r$ can be obtained with $\mathbf{V}\cdot\mathbf{E}_{r}$. The
amplified displacement $\delta_{w}$ at the CCD is much larger than
the initial shift $|\delta_{\pm}|$. Calculating the distribution of
$\mathbf{V}\cdot\mathbf{E}_{r}$ yields the amplified factor
$A_w=\delta_{w}/\delta_{+}$. Hence the amplified displacements at
the CCD are $A_{w}\delta_{\pm}$. It should be mentioned that the
amplified factor $A_{w}$ is not a constant, which verifies the
similar result of our previous work~\cite{Luo2011a}.
\begin{figure}
\includegraphics[width=6cm]{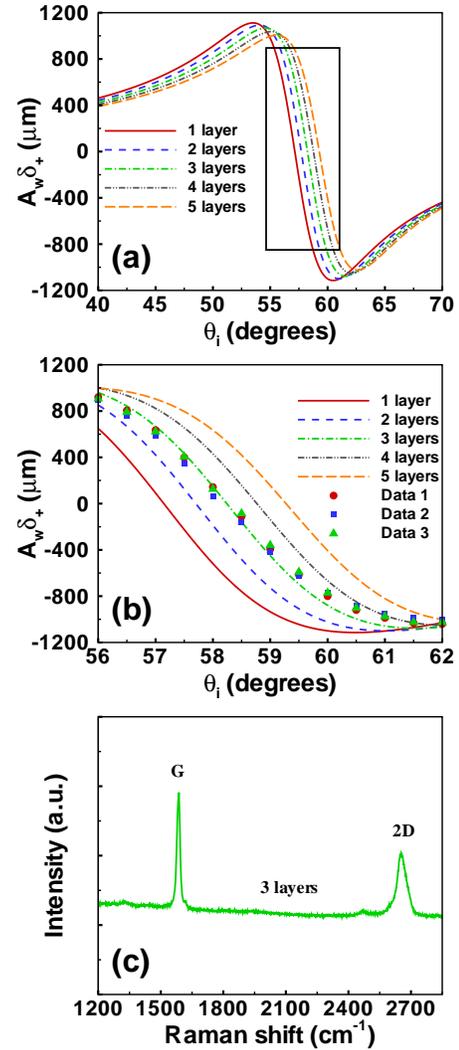}
\caption{\label{Fig4} (Color online) The theoretical and
experimental results of determining the layer numbers of graphene.
(a) represents the theoretical transverse displacements under the
condition of graphene layer numbers changing from one to five. Here
the refractive index of graphene is $3.0+1.149i$ at $633$ nm. (b)
describes the transverse shifts in the case of different incident
angles ranging from $56^{\circ}$ to $62^{\circ}$. The lines
represent the theoretical results. The circle, square and triangle
show the experimental data obtained from three different areas of
the graphene sample. (c) Raman reference data of the sample. }
\end{figure}

Now we focus our attention on identifying graphene layers. However,
there exists two unknown parameters (refractive index and layer
numbers of graphene) to be identified. Before identifying the
graphene layers, we need to choose the suitable refractive index
parameter of graphene. There has several measured values of the
refractive index of graphene reported
recently~\cite{Blake2007,Ni2007,Wang2008,Bruna2009}. Here we choose
one suitable refractive index according from the work of Bruna and
Borini~\cite{Bruna2009}. They concluded that the refractive index of
graphene in the visible range consists of real refractive index
(constant) and complex refractive index (depending on the
wavelength). Here the refractive index of graphene is about $3.0 +
1.149i$ at 633 nm. We first need to prove that this refractive index
is suitable for our graphene film. Our sample consists of graphene
films with two different layers: one layer, two layers. The graphene
films (made from ACS Material company) were first grown on
$25{\mu}m$ thick copper foil in a quartz tube furnace system using a
CVD method and then were transferred to the prism. The Raman spectra
of these two samples are shown in Fig.~\ref{Fig3}(a). We measure the
displacements of the SHE of light on the graphene film every
$2^{\circ}$ from $40^{\circ}$ to $70^{\circ}$ in the case of
horizontal polarization and the results are shown in
Fig.~\ref{Fig3}(b) and~\ref{Fig3}(c). It should be noted that the
quality of the material (graphene film) and the experimental
environment will affect the measurement. A group of experiment for
measuring the SHE of light at a pure air-prism interface was also
carried out for making a reference. We can find that the
experimental results fit well with the transverse displacement curve
calculated from the literature of Bruna~\cite{Bruna2009}. We can
obtain that the refractive index of graphene is really close to
$3.0+1.149i$ at $633$ nm. Therefore the SHE of light provides us an
alternative way for choosing the refractive index of graphene.

Using the suitable refractive index $n=3.0+1.149i$ at $633$ nm, we
can identify the layer numbers of an unknown graphene film with the
weak measurements. The experimental sample is also prepared with the
CVD method. It should be noted that, in our experimental condition,
we can not fabricate the sample with the precise layer numbers when
the graphene film has more than two layers. Because it would
unavoidably involve large technical errors. We just know the
approximate layer numbers ranges. Therefore we prepare a sample with
the possible layer numbers ranging from three to five layers. Our
aim is to determine the actual layer numbers of this graphene film.
Figure~\ref{Fig4} shows the theoretical and experimental results of
the graphene layer numbers determination. From Fig.~\ref{Fig4}(a),
we find that it is hard to distinguish the transverse shifts of the
different graphene layer numbers from $40^{\circ}$ to $70^{\circ}$.
Thence we measure the transverse displacements in a small range of
incident angle (from $56^{\circ}$ to $62^{\circ}$) to obtain a
desired results [Fig.~\ref{Fig4}(b)]. To avoid the influence of
impurities and other surface quality factors of graphene film, we
carried out the experiments for three different areas of the
graphene sample. From the experimental results, we can conclude that
the actual layer numbers of the film is three. To further confirm
our results, we also add some reference data via measuring the Raman
spectra of the sample as shown in Fig.~\ref{Fig4}(c). Therefore the
SHE of light can become a useful metrological tool for
characterizing the layer numbers of graphene.

In conclusion, we have presented a simple and convenient method for
determining the layer numbers of graphene. Firstly, we use the SHE
of light to choose the suitable refractive index of graphene
obtaining from the corresponding literature. And then, with the
suitable refractive index $n=3.0+1.149i$ at $633$ nm, the layer
numbers of an unknown graphene film can be detected with desired
precise. So combining the SHE of light with the other techniques
such as atomic force microscopy and Raman spectroscopy can improve
the graphene layers identification accuracy, which is important for
the future graphene research.

One of the authors (X. Z.) would like to thank Dr. Sosan Cheon for
helpful discussions. This research was partially supported by the
National Natural Science Foundation of China (Grants Nos. 61025024
and 11274106) and Hunan Provincial Natural Science Foundation of
China (Grant No. 12JJ7005).

\end{document}